\documentclass[11pt, oneside]{mnthesis}
\usepackage{epsfig,epic,eepic,units}
\usepackage{hyperref}
\usepackage{url}
\usepackage{longtable}
\usepackage{mathrsfs}
\usepackage{multirow}
\usepackage{bigstrut}
\usepackage{amssymb}
\usepackage{graphicx}
\graphicspath{ {./figs/} }

\begin{document}
\bibliographystyle{hunsrt} 

\ms 

\title{\bf Adaptive Domain Generalization for Digital Pathology Images}
\author{Andrew John Walker}
\degree{MASTER OF SCIENCE}
\director{Prof. Ju Sun} 

\submissionmonth{May} 
\submissionyear{2022} 

\abstract{

In AI-based histopathology, domain shifts are common and well-studied. However, this research focuses on stain and scanner variations, which do not show the full picture-- shifts may be combinations of other shifts, or "invisible" shifts that are not obvious but still damage performance of machine learning models. Furthermore, it is important for models to generalize to these shifts without expensive or scarce annotations, especially in the histopathology space and if wanting to deploy models on a larger scale. Thus, there is a need for "reactive" domain generalization techniques: ones that adapt to domain shifts at test-time rather than requiring predictions of or examples of the shifts at training time. We conduct a literature review and introduce techniques that react to domain shifts rather than requiring a prediction of them in advance. We investigate test time training, a technique for domain generalization that adapts model parameters at test-time through optimization of a secondary self-supervised task.

}
\copyrightpage 
\acknowledgements{

This work was conducted in partnership with PathAI. I would like to express my gracious thanks to the company and my mentors: Dinkar Juyal, Chintan Shah, Syed Ashar Javed, Harsha Pokkalla, and Adi Prakash. I have learned so much from you all and appreciate all your support and mentorship.

I would like to thank Prof. Ju Sun, my advisor, for his unwavering support of my research. Thank you also to all the UMN GLOVEX lab members for many stimulating conversations and their camaraderie. Finally, I would like to thank my defense committee members: Prof. Ju Sun, Prof. Daniel Boley, and Prof. Bharat Thyagarajan; thank you for your expertise. 

}
\dedication{Dedicated to my family, who have always supported me and my academic pursuits. }


\beforepreface 

\figurespage
\tablespage

\afterpreface            


\chapter{Introduction}
\label{intro_chapter}

It is a key assumption for theoretical performance guarantees of machine learning models that the test data is drawn from the same distribution as the training data. Domain shift is the term used for when the test data is drawn from a different distribution than the training data. Domain shift appears naturally and can severely negatively impact a model's expected performance.

Domain shift is a relatively new area of study in machine learning-- as models are increasingly deployed "in the real world", generalization to new test sets becomes more important. Instead of eking out a slightly higher accuracy on the test set (which in the standard machine learning setting has the same distribution as the training set), effectively overfitting to the test distribution, it has become more relevant to deploy robustly to unforeseen settings.

This idea of domain generalization has become more popular in recent years, and many special datasets and challenges targeting the domain shift issue have arisen \cite{wilds, imagenet_c, imagenet_a, imagenet_r, imagenet_v2, cifar_10_1, colored_mnist}. Domain generalization becomes especially important in histopathology, as acquiring new annotations for new domains is prohibitively expensive and domain shifts are incredibly common: different models of slide scanner or slightly different staining procedures at different labsites are inconsistent enough to be considered different domains \cite{learning_domain_invariant_representations_of_histological_images, quantifying_the_effects_of_da_and_color_norm_in_pathology}. 

This thesis introduces domain generalization as a field and domain shift in histopathology, defines our specific problem setting, presents a review of existing techniques, and describes the selected technique of Test Time Training. It details experiments regarding the implementation of Test Time Training. This thesis concludes by revisiting the need for domain generalization in histopathology.

\chapter{Preliminaries}
\label{preliminaries_chapter}

This chapter introduces some necessary prerequisite knowledge about deep learning as a foundation for the rest of the thesis.

\section{Deep Learning}
Machine learning is a problem solving technique that uses algorithms that automatically learn relationships through data. The algorithm can also be called a model, and the aforementioned process can be called "training." A subset of machine learning is deep learning, which is a name for the new wave of artificial neural networks (ANNs), a type of machine learning model, which are increasingly "deep" and computationally intensive. New improvements in computational power and data availability have led to deep learning becoming one of the most popular approaches in the ongoing third renaissance of deep learning, paving the way to solve difficult problems in fields such as medicine, climate, automation, and many more. Deep neural networks (DNNs) have seen incredible success in image processing, often outperforming human ability. This performance is only outstanding in some sense: Although DNNs may achieve higher accuracy in specific sterilized settings, this superhuman performance generally requires deployment of the network in an environment very similar to that it was trained in. DNNs are surprisingly brittle, failing when there are small variations between test data and training data that humans are robust to, e.g. blurring, different cropping, or static noise. The elimination of this brittleness in DNNs is the key focus of this thesis, and will be expanded upon in \ref{background:domain_shift_robustness_in_compvis}.

\begin{figure}
\centering
\begin{subfigure}{\textwidth}
  \centering
  \includegraphics[width=0.5\linewidth]{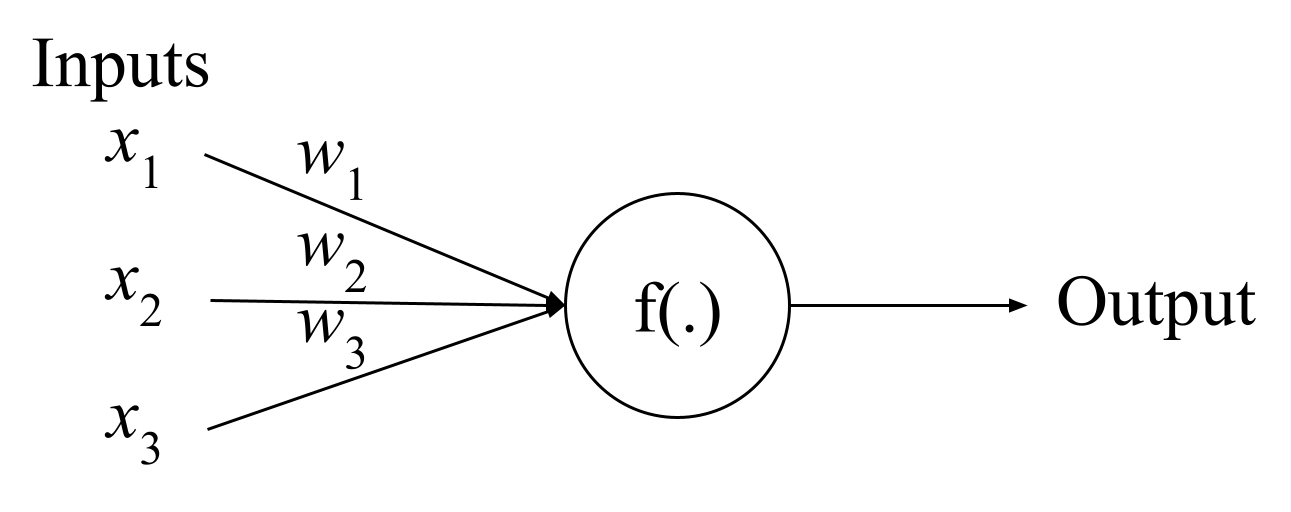}
  \caption[The Perceptron]{The perceptron. It takes as input some signal (x) from either the input data or a previous perceptron. It weights these signals according to some learned parameters (w) before performing an operation on them, generally a summation followed by an activation function (e.g. sigmoid, ReLU, etc.). Then, this output can be fed to another perceptron, or can be used as the desired output of the neural network.}
  \label{fig:perceptron}
\end{subfigure}

\begin{subfigure}{\textwidth}
  \centering
  \includegraphics[width=0.33\linewidth]{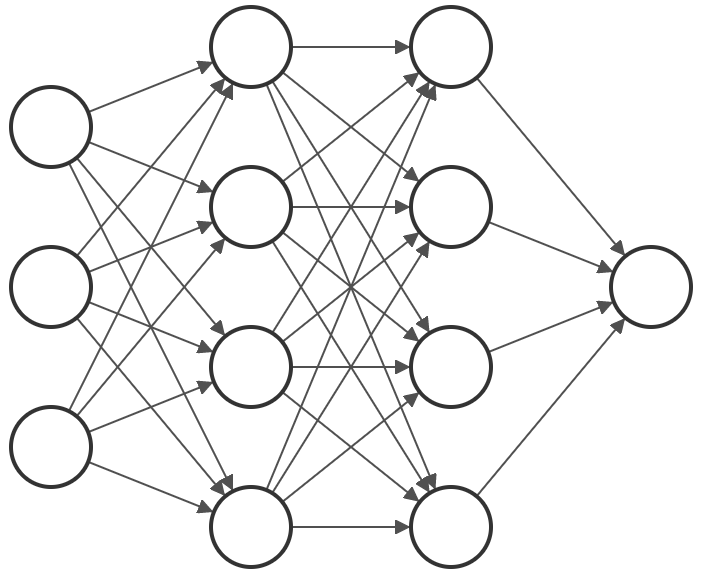}
  \caption{Multiple layers of perceptrons come together to form an artifical neural network.}
  \label{fig:ann}
\end{subfigure}

\caption[Structure of an Artificial Neural Network]{Structure of an artificial neural network.}
\label{fig:ann_structure}
\end{figure}

The deep neural network is often analogized to biological neural networks, i.e. the human brain or human visual system, although it is only loosely inspired in structure and not in learning technique. The base unit of the deep neural network is called the "perceptron" after the neuron (Figure \ref{fig:perceptron}).

A perceptron is usually found in parallel with others in a "layer" to provide specialization among them, and these "layers" of perceptrons are strung together in a fully-connected manner (i.e. each perceptron in one layer is connected to each perceptron in the next layer). This is a fully connected network, but other schemes have been proposed with different connection patterns. One popular architecture for image processing is the convolutional network, which pares down the number of connections from the fully-connected in such a way as to improve performance on translational variance. 

To train an Artificial Neural Network, a loss function is designed to have a lower value at more appropriate outputs, creating a loss landscape that is locally optimizable through optimization methods like stochastic gradient descent or Adam. 

As ANNs become deeper and wider, they become better at interpolating training data, able to learn even randomly assigned labels. This is good for test data that is similar to the training data, but does not guarantee much for data dissimilar to the training data, and this is the reason for their brittleness to domain shifts. This concludes the preliminary introduction of deep learning, and in chapter \ref{background_chapter}, the idea of domain shift adaptation is introduced. The next preliminary chapter briefly introduces some differences between image processing in digital pathology and the standard setting.

\section{Digital Pathology vs Natural Images}
Histopathology is the diagnosis of disease via visual examination of tissue under a microscope \cite{DL_in_histo_a_review}. Whole Slide Images (WSIs) are digital scans of the entire tissue slide, and they have helped open digital and AI analysis of the tissue samples. 

The most popular academic setting for computer vision is natural image classification or segmentation. Histopathology images are a unique domain apart from natural images, and provide their own set of idiosyncracies. 

Natural images are generally small, around 250x250 pixels, whereas WSIs can be gigapixel size. Also, WSIs can be sampled at varying magnification levels. Natural images often contain a small number of distinct objects, while WSIs include many individual cells and anatomical structures. While annotations for natural images are abundant, annotations are scarce and expensive in histopathology. And, WSIs have unique artifacts not seen in natural images and different natural domain shifts than are seen in natural images.

\chapter{Background}
\label{background_chapter}
\section{Domain Shift Robustness in Computer Vision}
\label{background:domain_shift_robustness_in_compvis}

\paragraph{Parallels to Generalization and Robustness}
The idea of domain shift is closely related to model generalization and robustness. Generalization refers to a model's performance on test data versus training data. A model is robust if small perturbations to a training point result in small changes in the output \cite{robustness_and_generalization}. Note that these perturbations are not qualified in any way, that is, in the robustness setting, there are no stipulations on what makes a perturbation "legal". A model must generalize well to all \emph{small} perturbations. This is why the standard machine learning paradigm requires test data and training data to be drawn from the same domain. 

However, domain shift can result in \emph{large} perturbations to features in some \emph{patterned} manner. A domain shift has some sense of collectiveness: all data points in a domain are expected to be different in some meaningful and summarizable manner from all points in a different domain, whereas perturbations in the generalization setting are more sporadic: there is no sense of collection or meaning. These two properties of domain shifts (collectiveness and meaningfulness) can be exploited to achieve domain generalization more easily than normal generalization techniques may allow. 

Domain shift is a special case of model generalization-- we care that a model becomes robust to domain-changing perturbations. Furthermore, many techniques for domain generalization draw inspiration from or are even shared with generalization and robustness. Because of this, we introduce some relevant generalization and robustness concepts. 

In model training, the training error of a model is directly optimized to some local minimum. However, the use of a model is not to memorize training data, but to be readily deployed to new data. That is, a model is only valuable if it is able to achieve good test error. Generalization refers to a model's ability to perform well (as well as it does on training data) on unseen test data \cite{deep_learning_book}. Domain generalization refers to a model's ability to perform well on test data that is subject to some domain shift.

Because a model is optimized on the training data, the training error is expected to be lower than or equal to the test error. That is, a small gap between the train error and test error is expected. However, when a model is able to achieve low training error but that gap is too large, the model has "overfit" the training data. This means the model has learned some spurious correlations between the training features and the training outputs that do not appear in the test data. This is a case of bad generalization, and is also a relevant problem in domain generalization.

Taking a robustness perspective, we can imagine generating pseudo-test samples by taking train samples and applying some reasonable constrained perturbations to them \cite{adversarial_ml_tutorial}. In this thought experiment, reasonable and constrained means we add a noise term to each pixel that is small enough to be imperceptible to humans. This is the $\ell_\infty$ ball, where the perturbation is described by $$\Delta = \{\delta : ||\delta||_\infty \leq \epsilon\}$$
Imagine that we are acting adversarially, or hope to cause the most damage to the model's prediction. We have control over what perturbation to apply, so long as it exists in $\Delta$. We hope to select the worst-case perturbation, or an "adversarial example": the one that most damages the model's prediction. This is termed an "adversarial attack". Adversarial training is a technique to improve robustness on $\Delta$ perturbations by training on adversarial examples. If we manage to train our model so that it is robust to any $\Delta$ perturbation on the input space, this means that it is generalizes well to the perturbation space. 

To bridge the gap between generalization and domain generalization: if we create a new set of perturbations $\Delta_{\text{shift}}$ that encompass any perturbations that arise from a domain shift, we can imagine replicating this procedure to create a model that generalizes well to domain shift. This is one perspective by which to see domain shift, and shows how it is analogous to standard generalization-- they are often treated similarly in the literature. 







\paragraph{Back to Domain Shift}
To formalize domain shift, we first introduce notation. A domain is the combination of input space \calX, output space \calY, and the probability distribution $p$ mapping between the two. Because the input space and output space are the same in our situation, we simply denote a domain with its probability distribution $p$. Here, the source domain is \psrc and the target domain is \ptar.
Some examples: we can refer to the source domain's joint distribution with $\psrc(x,y)$, or the target domain's class conditional distribution with $\ptar(x | y)$. 

Different types of domain shift exist, and names have been given to some special cases: prior shift, concept shift, and covariate shift
\cite{technical_report_intro_domain_adaptation}. Of course, natural domain shifts can be complicated mixes of these. These three special cases can all be understood by observing two decompositions of the domain's joint probability distribution: $p(x,y)=p(x|y)p(y)$ and $p(x,y)=p(y|x)p(x)$.
Prior shift occurs when prior distributions of the classes are not the same, $\psrc(y) \neq \ptar(y)$, but the conditional distributions are still equal, $\psrc(x|y) = \ptar(x|y)$. In the histopathology setting, this may occur when the model is inadvertently trained on a population with a much lower background disease rate than it is deployed on. Although the disease may exhibit the same features in the input space \calX in both domains, the frequency of disease in the target domain is much higher. This can be serious, as it may bias the model to provide false negative results, but it can be addressed by class weighting during training and selecting a model with good precision and recall instead of solely based on accuracy. We do not focus on this type of domain shift in this project. 

Concept shift occurs when the posterior distribution of the classes are not the same between train and test sets, $\psrc(y | x) \neq \ptar(y | x)$, but the $x$ marginal distribution remains constant $\psrc(x) = \psrc(y)$. This may occur when the meaning of the class labels changes between domains -- for instance, if some medium-severe cases of disease were binarized as healthy in the training data but in deployment we wish to classify them as diseased instead. Observe that the feature distribution remains constant, but that the class definitions, and thus the class posteriors have shifted in spite of this. To adapt in this setting, labeled target data is required. For example, this scenario can be addressed by transfer learning. We do not focus on this type of domain shift in this project. 

Finally, covariate shift is where the posterior distribution of the classes remains the same, $\psrc(y|x) = \ptar(y|x)$, but the distribution of the features differ between the training and test sets, $\psrc(x) \neq \ptar(x)$. Actually, this term has come to mean something slightly different, and covariate shift research does not require $\psrc(y|x) = \ptar(y|x)$. 

Covariate shift may occur when the feature generating step has different noise in both domains. An example from computational histopathology is where different WSI scanners might provide different colorations, resulting in different feature distributions. This is a basic example, but there are many other sources of potential variation, some more complex. We focus on covariate shift because it is the most pertinent to the business needs; although the other shifts can occur as well, they are generally easier to deal with. Details on covariate shift in histopathology imaging will be in section \ref{covariate_shift_in_computational_histopathology}.

\begin{figure}
    \centering
    \includegraphics[width=\textwidth]{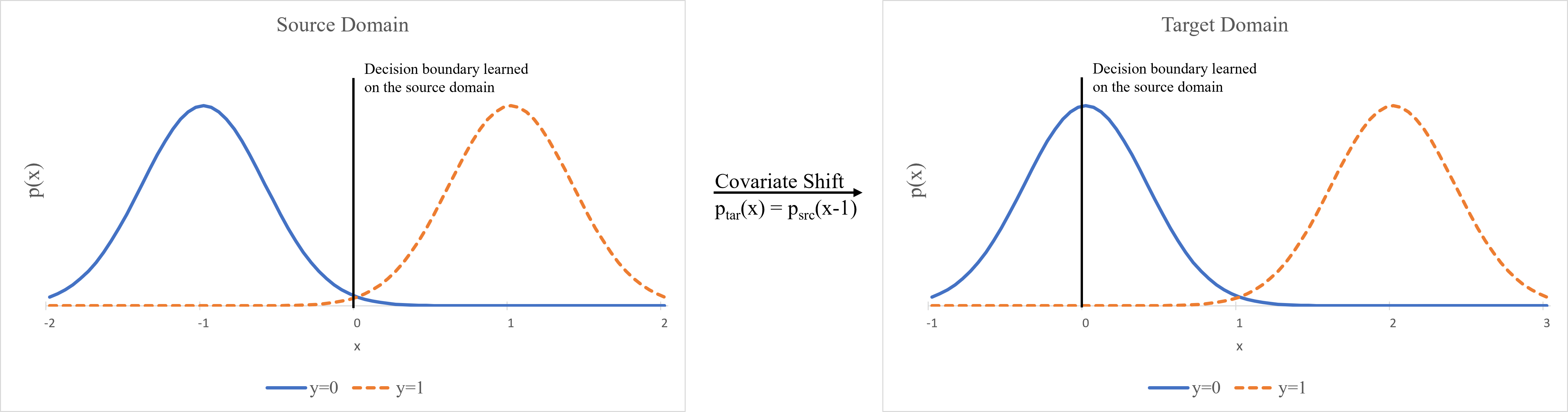}
    \caption[Toy Example Showing how Covariate Shift can Damage Model Performance]{Toy example showing how covariate shift can damage model performance. The covariate shift on the target domain results in the decision boundary learned on the source domain to be erroneous.}
    \label{fig:covariate_shift_ex}
\end{figure}

\section{Covariate Shift in Computational Histopathology}
\label{covariate_shift_in_computational_histopathology}

\subsection{Sources of Covariate Shift in the Data Collection Pipeline}
\begin{figure}
    \centering
    \includegraphics[width=\textwidth]{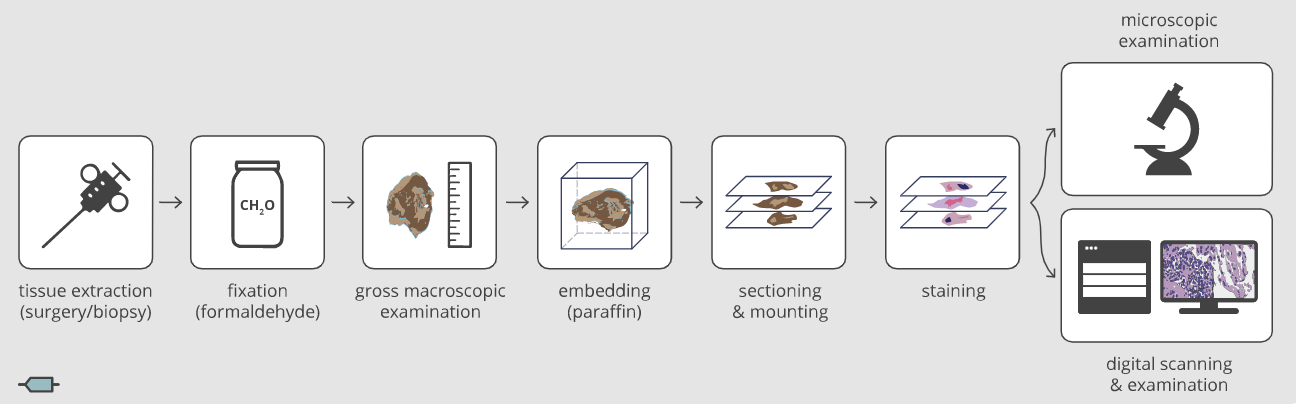}
    \caption[Whole Slide Image Creation Pipeline]{Whole slide image creation pipeline. Image from PathAI.}
    \label{fig:wsi_creation_pipeline}
\end{figure}

\begin{figure}
    \centering
    \includegraphics[width=\textwidth]{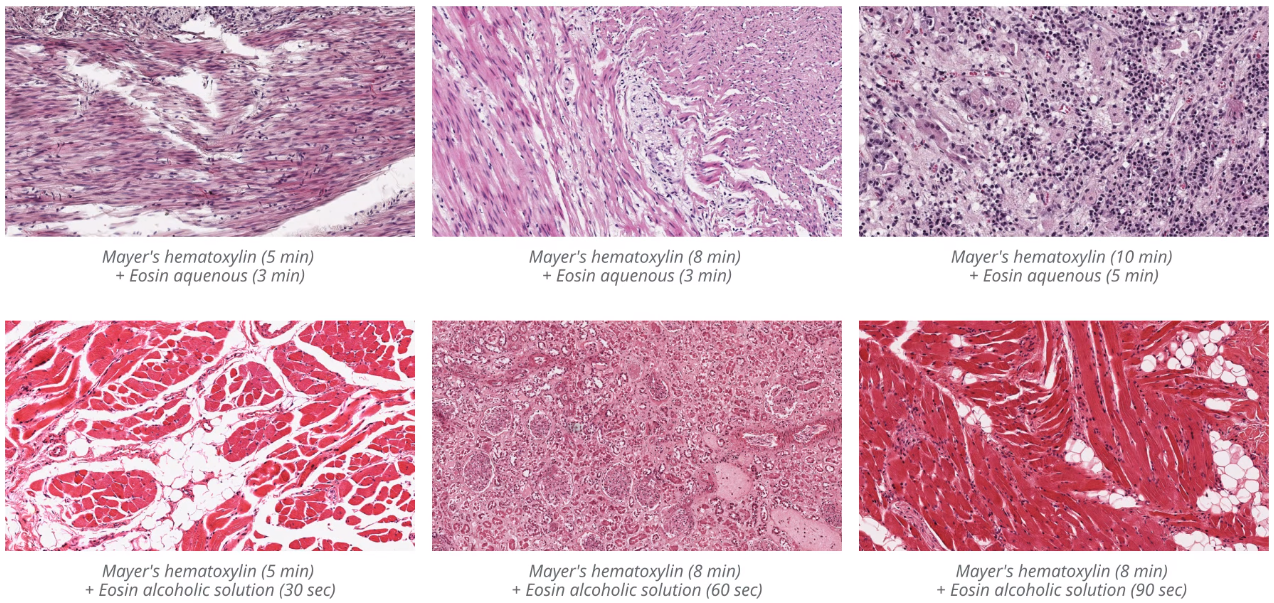}
    \caption[Examples Showing how Staining Protocol Affects Slide Appearance]{Example showing how staining protocol can affect slide appearance. From BioVitrium's video: \href{https://www.youtube.com/watch?v=1tqA_gWUjkM}{``Histological staining: hematoxylin \& eosin"}. 
    }
    \label{fig:stain_variation}
\end{figure}

\begin{figure}
    \centering
    \includegraphics[width=\textwidth]{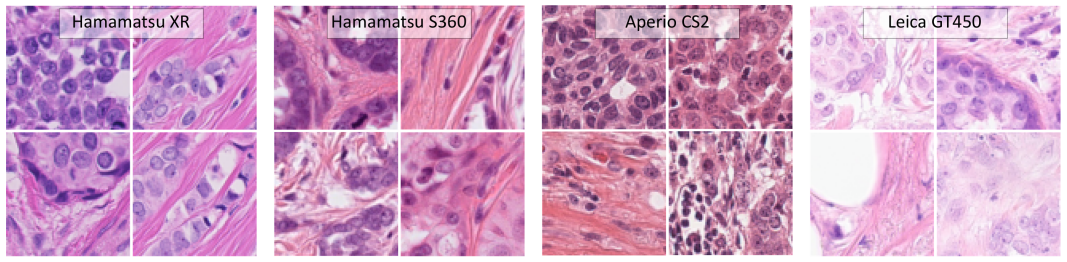}
    \caption[Variations Due to Scanner Type]{Variations due to scanner type. From \cite{aubreville_scanner_domains}.}
    \label{fig:scanner_variation}
\end{figure}

The data collection pipeline for these models has many steps that can introduce variations that constitute the possible domain shift seen at test time \cite{dl_in_dp_survey}. The pipeline (Figure \ref{fig:wsi_creation_pipeline}) begins with tissue extraction, usually core needle biopsy or fine-needle aspiration. Different collection types can impact the appearance of the WSI. The samples are chemically fixed, or preserved, with varying media. The fixed tissue sample is then embedded in a block, which can be of varying material. It is then sectioned and mounted on glass slides, but the width of sections can vary. At this point, the tissue sections are still transparent. The slices must be chemically stained to selectively pigment different structures in the anatomy. Common stains are Hematoxylin and Eosin, for example. All variations up to this point can impact the appearance of the WSI. Even within the same staining type, variations in staining protocol can impact the end appearance of the WSI (Figure \ref{fig:stain_variation}). There is no universal staining procedure, because human pathologists have different preferences. Finally, the stained slides are scanned by a slide scanner, which may have varying degrees of human autonomy, and which different models have different color representations (e.g. color, contrast, saturation) and optical parameters (e.g. depth of field, blurring, sharpness) \cite{mitosis_midog}. These are a source of variation as well (Figure \ref{fig:scanner_variation}). The high variability in WSI images has an adverse impact on machine learning algorithms for histopathology \cite{quantifying_the_effects_of_da_and_color_norm_in_pathology}. 

\begin{figure}
    \centering
    \includegraphics[width=0.5\textwidth]{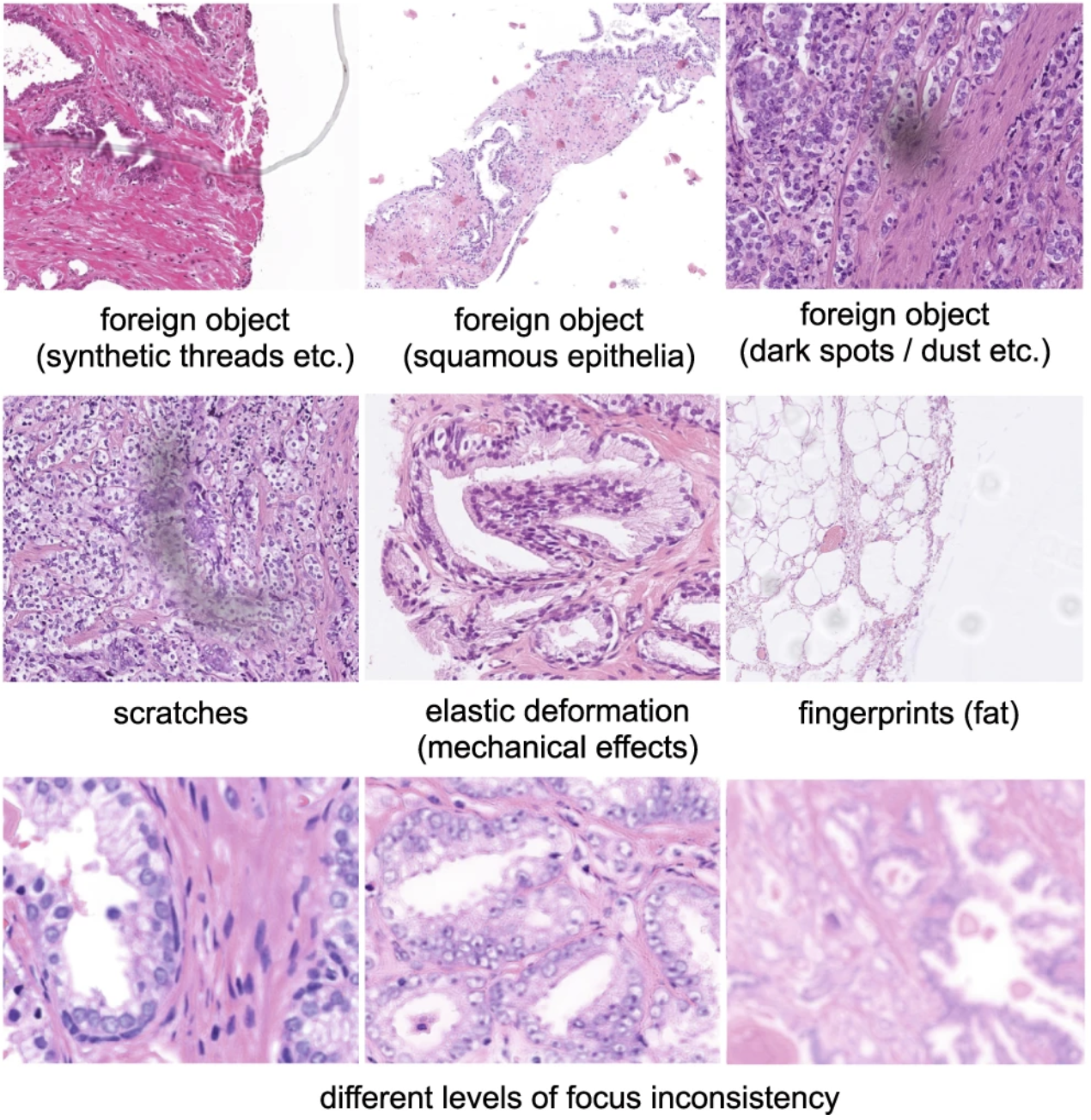}
    \caption[Examples of Physical Artifacts Appearing in Whole Slide Images]{Examples of physical artifacts appearing in whole slide images. From \cite{examples_of_artifacts}.}
    \label{fig:artifacts}
\end{figure}

Many types of artifact arise during these steps too, which are a type of covariate shift as well, shown in Figure \ref{fig:artifacts}. However, as these are easy to remove ad-hoc from inference, they are not the point of focus for this project, which is global shifts like stain variations, scanner variations, complex combinations of these, and even intangible shifts that are indescribable by humans, such as those appearing from different labs. 



\subsection{Different "Flavors" of Domain Generalization}
Work in the area of domain generalization has become increasingly popular as models are applied in the real world with domain shift (e.g. self-driving cars in inclement weather conditions), and accuracy on a "sterilized" test-set is no longer the most important metric. As domain generalization research has expanded, different researchers have found interest in exploring different "flavors" of domain generalization, each allowing access to different amounts of test data and requiring different assumptions of the data shift.

The general form of domain generalization involves M source domains and a target domain. In domain generalization, we hope to learn a general function $h: \calX \ra \calY$ from all M source domains and achieve good performance on any arbitrary test domain (but approximated by the one target domain) \cite{generalizing_to_unseen_domains_survey}.

Multi-task learning, or joint training, is when a model is trained on $\calX$ with multiple tasks $\calY^i$ \cite{multitask_learning}. Sometimes, the tasks may all be of interest, as in the case of segmentation and classification of tissue, or sometimes, only one of the tasks may be of interest. In the latter case, the important task is termed the main or primary task, and the other task(s) are termed the auxiliary or secondary task(s). Multi-task learning can improve robustness by sharing representations across all its tasks. Joint training can be used to improve domain generalization, and this is shown in the Test Time Training paper \cite{ttt}.

Domain Adaptation is identical to domain generalization with the caveat that we can access unlabeled samples from the test data $x_i \sim \ptar(x)$ during training.

There are other settings as well, but these are the most salient in building a foundation to understand Test Time Training.

        
\subsection{Our Application Setting}
\label{our_setting}
We do not conduct vanilla domain generalization, we explore a unique setting described in this section. The use case for domain shift robustness in digital pathology is to allow existing models to be deployed in new settings with protection against domain shift. This would be an alternative to transfer learning or training a new model from scratch, which both require expensive annotations. Additionally, in some cases, these annotations cannot be acquired: if the model is being deployed behind closed doors, data privacy may be important; for end users, they may not have the resources to provide annotations for new domains that appear. This is why it is desirable to have a model that generalizes well to new domains. The technique must also not have any negative effect when a domain shift does not exist, or there must be some way of turning off the technique if there is no domain shift.

We are interested in the setting where we can train on a single domain and generalize well to an arbitrary test domain (approximated by a single test domain). We also assume we do not get access to test data during model training, or access to train data during testing. This seemingly hyper-strict setting is of interest because there exist other good-performing techniques for a standard domain-generalization problem in which we have multiple training domains (e.g. meta-learning, enforcing domain invariant representations, pretraining on multiple domains) or some access to the test domain during training (e.g. transfer learning, domain adaptation techniques). Furthermore, it may be infeasible to expect access to the train data during test-time due to data privacy.

In the case of redeploying an existing model in a new project or lab, our setting allows for a domain to consist of multiple patients. The test domain in this case is defined by sharing the same data generating pipeline (i.e. project or lab WSI pipelines). This is a valuable and realistic domain shift setting. We can also consider a stricter setting where we are only allowed access to a single patient at a time. This setting arises when a model is deployed on a patient-by-patient basis, such as in a clinical setting. Because even a single WSI and a single patient may have some domain shift with respect to another, this approach provides the most flexibility with the trade-off of a large amount of variance. 

These models are medical devices, and FDA-approved medical devices are subject to specific repeatability and fairness regulations. The model must not provide different results depending on the order of patients in inference.

For some other miscellaneous requirements: there is no requirement for inference latency. Real-time inference is not a requirement for this model like it may be for real-time video models.

\subsection{Types of Domain Generalization Techniques}
One taxonomy outlines three families of domain generalization techniques \cite{generalizing_to_unseen_domains_survey}. These are: data manipulation, which manipulates the input data at training time to encourage domain invariance (e.g. data augmentation and data generation); domain-invariant representation learning, which specifically seeks domain invariance (e.g. adversarial training, feature alignment between domains, invariant risk minimization, or feature disentanglement); learning strategy, which modifies the learning strategy to improve generalization (e.g. self-supervised learning, meta-learning) 
We now focus on specific techniques from these three families that are relevant to understanding the experimental methods of this project.

\subsubsection{Data Augmentation}
Data augmentation is the now-"ubiquitous" \cite{albumentations} practice of applying label-preserving transformations to input data before using it as training examples for the model \cite{data_augmentation_survey}. The goal behind this is to provide a more-varied training set to ensure robustness. 

\begin{figure}
    \centering
    \includegraphics[width=0.5\textwidth]{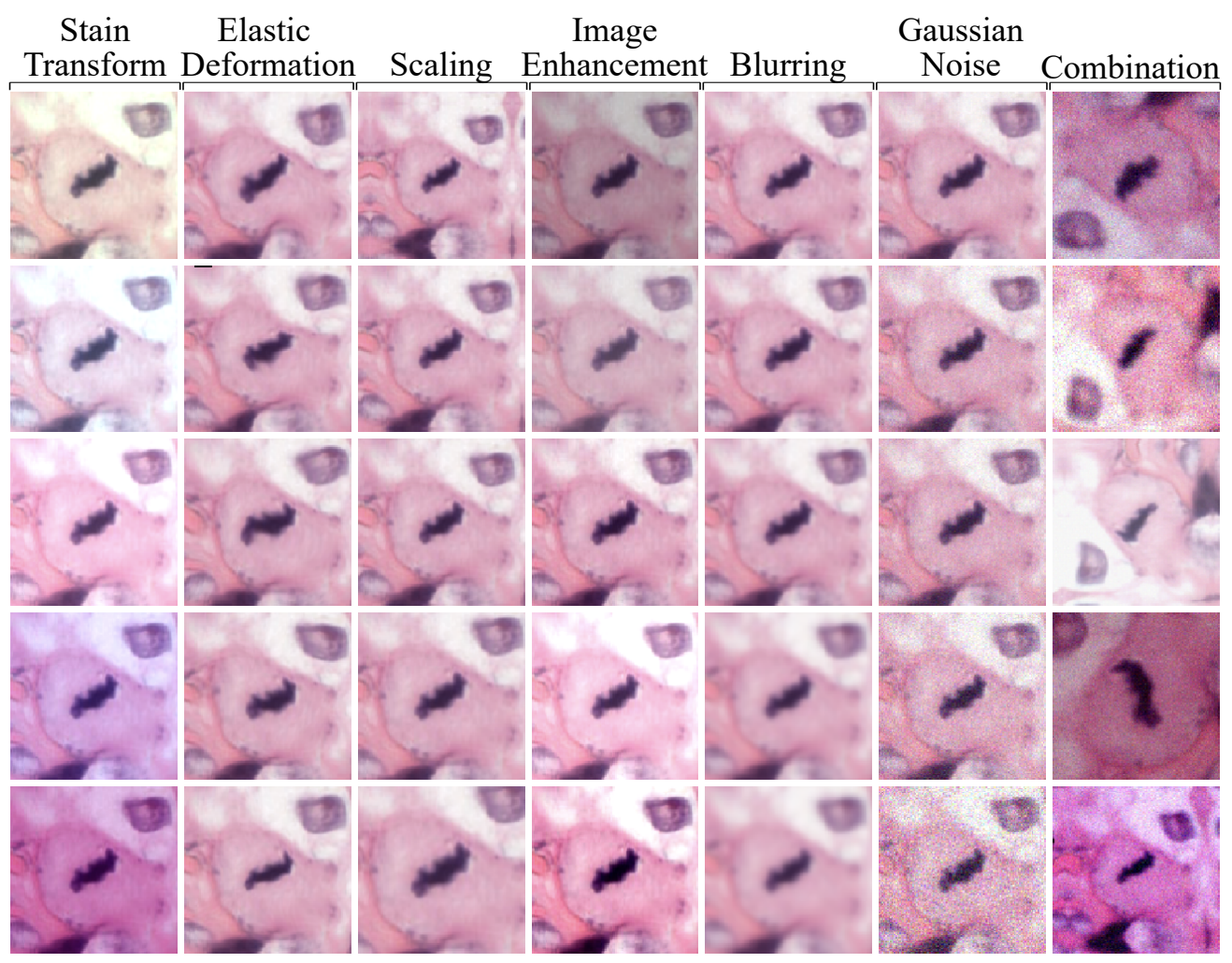}
    \caption[Examples of Histopathology-Inspired Data Augmentation]{Examples of histopathology-inspired data augmentation. From \cite{stain_augmentation_mitosis_detection}.}
    \label{fig:histo_augs}
\end{figure}

These transformations can be geometric transformations, like flipping and cropping, or photometric shifts, or noise injection. Some histopathology-inspired augmentations exist, including elastic deformation to mimic differently-shaped anatomical structures, gaussian blurring to mimic out-of-focus scanning artifacts, jpeg compression to simulate image compression artifacts, and H\&E stain augmentation to simulate staining variations by disentangling the hematoxylin and eosin color channels and independently perturbing them \cite{stain_augmentation_mitosis_detection}. 
Figure \ref{fig:histo_augs} provides some examples of these histopathology-inspired transformations. 


The benefit of data augmentation is twofold: it can increase dataset size and improve dataset diversity \cite{albumentations}. The size of the dataset is increased by allowing each original data point to become many more. Increasing the size of the dataset can reduce overfitting to the training set. The improvement of dataset diversity is more relevant to dmoain generalization, however. Using augmentations exposes the model to corruptions, thus improving robustness to those types of corruptions. Selecting augmentations for attributes that change across domains can encourage domain shift robustness.

Using data augmentations to supplement training data in the normal training paradigm can be helpful, but augmentations have also found use in contrastive self-supervised learning, which will be discussed partway through the next section.

\subsubsection{Self-Supervised Learning}
\label{self_supervised_learning}
Self-supervised learning (SSL) is an unsupervised learning method similar to traditional supervised learning, except that the labels are created from the data itself, for so-called "pretext tasks". Self supervised learning is used when there is an abundance of unlabeled data and scarce labeled data. In this case, SSL is used to pretrain the model with the unlabeled data, learning adequate representations along the way; this model is then fine-tuned to the target task with the scarce labeled data \cite{sun_unsupervised_da_through_self_sup}.

\begin{figure}
    \centering
    \includegraphics[width=0.4\textwidth]{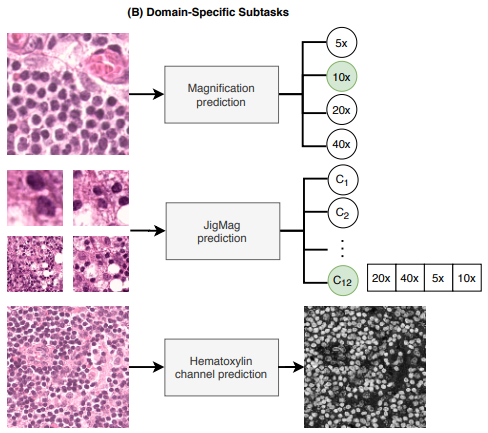}
    \caption[Histopathology-Inspired Self-Supervision Tasks]{Histopathology-inspired self-supervision tasks. From \cite{selfpath_jigmag}.}
    \label{fig:selfpath_tasks}
\end{figure}

Many pretext tasks have been proposed, including human-designed tasks such as image rotation prediction or solving image jigsaw puzzles, and there are even some histopathology-informed tasks, like slide magnification prediction (See Aside \ref{RSP_aside}, as this is important to our methods) and stain separation, shown in Figure \ref{fig:selfpath_tasks}\cite{selfpath_jigmag, martel_ssl}. Again, the goal of such pretext tasks is to jump-start the model in learning useful representations of the data-- for example, to properly predict the magnification of a histopathology slide image, the network might learn what different anatomical structures look like at different magnifications. These learned representations would then be useful in fine-tuning to a main task.

\begin{aside}
    \label{RSP_aside}
    \textbf{Aside \ref{RSP_aside}: Resolution Sequence Prediction (RSP)} \cite{martel_ssl} is a histopathology-inspired self-supervised task in which the model must predict the optical magnification of three concentric WSI samples. We use 40x (0.25mpp), 20x (0.5mpp), and 10x (1mpp). The model knows there is one and only one image at each possible magnification, and can reference all the images at once. This is a $3! = 6$ class problem, as each patch can take one of 3 optical magnification levels without repetition. We use this task in our experiments. This task is very similar to the JigMag task found in \cite{selfpath_jigmag}. 
    
    \begin{center}
      \includegraphics[width=0.5\textwidth]{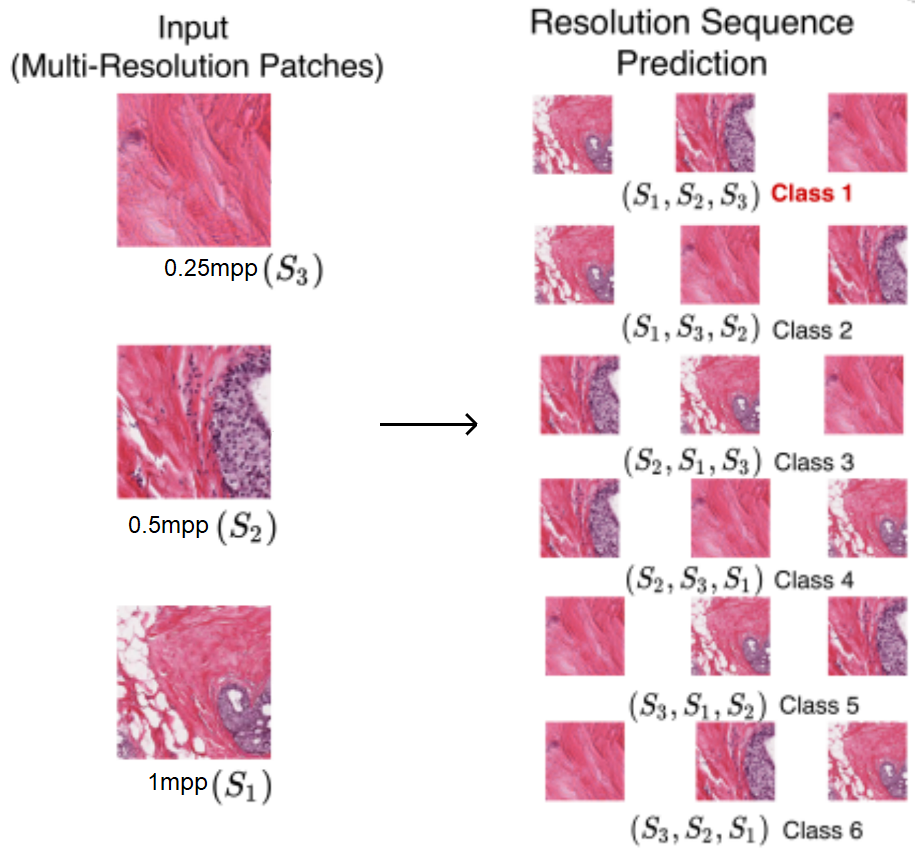}
      \captionof{figure}[The Resolution Sequence Prediction Task]{The resolution sequence prediction (RSP) task. Adapted from \cite{martel_ssl}.}
      \label{fig:rsp_task}
    \end{center}
\end{aside}

Other popular pretext tasks are contrastive learning based tasks. The core of contrastive learning is to learn by comparing examples, learning to pull representations of similar examples (i.e. positive examples) together and push representations of dissimilar examples (i.e. negative examples) apart \cite{contrastive_learning_review}. Some techniques pair contrastive learning with image augmentation, creating a self-supervised contrastive model where positive examples can be automatically generated by applying different augmentations to the same base image, and negative examples start from different base images \cite{simclr} (See Aside \ref{SimCLR_aside}, as this is important to our methods).

Self-supervision's intended use is to use unlabeled data to improve generalizability, not specifically domain generalization, but it has nonetheless found use in this application. For example, if the unlabeled data comes from many domains, and a domain invariant pretext task is chosen, then representations learned by self-supervised learning would be well adapted to the provided domains and hopefully adapt well to new target domains \cite{sun_unsupervised_da_through_self_sup}. Alternatively, in a joint training setup, self-supervision plus supervision on the source domain alone has been shown to improve robustness and therefore domain generalizability to some extent \cite{hendrycks_joint_ssl_robustness}.

\begin{aside}
    \label{SimCLR_aside}
    \textbf{Aside \ref{SimCLR_aside}: "A Simple Framework for Contrastive Learning of Visual Representations" (SimCLR)} \cite{simclr} is a self-supervised contrastive learning task that we use in our method. It uses simple image augmentations on a set of inputs, creating two augmented views of each image. The task is to pull together representations of the same image with different augmentations and push apart representations of different images.
    
    
    \begin{center}
      \includegraphics[width=0.5\textwidth]{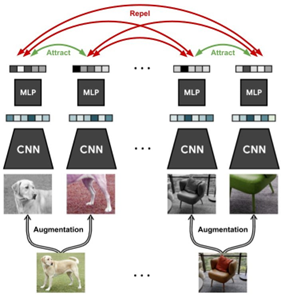}
      \captionof{figure}[The SimCLR Task]{The SimCLR task. From \cite{simclr}.}
      \label{fig:simclr_blog}
    \end{center}
\end{aside}

\subsubsection{Feature Alignment}
\begin{figure}[h!]
    \centering
    \includegraphics[width=0.5\textwidth]{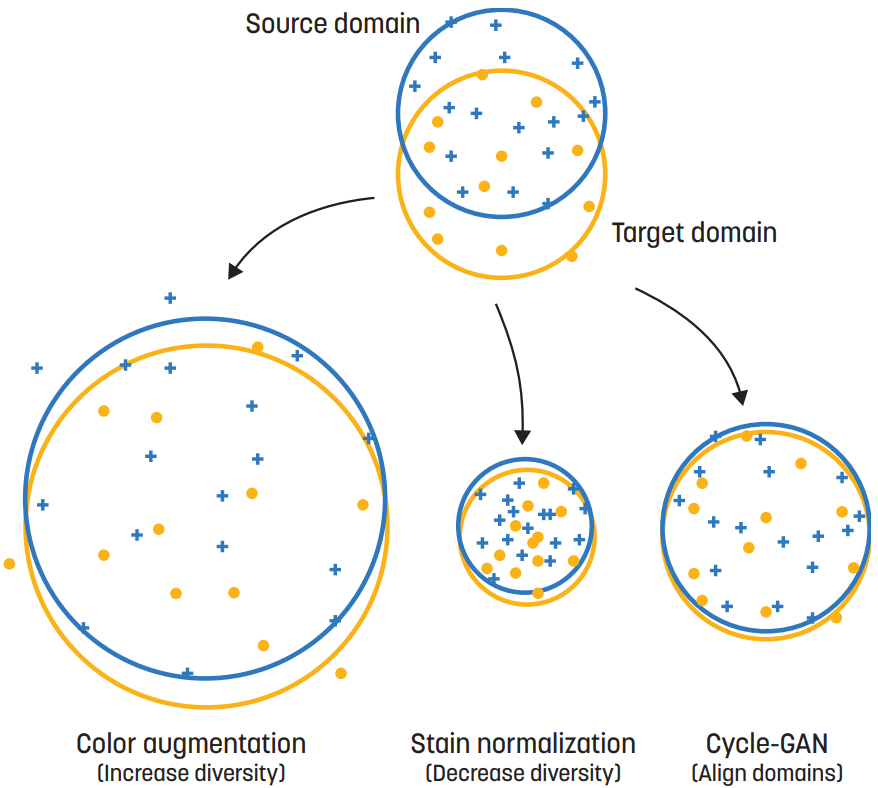}
    \caption[Augmentation vs Alignment in Generalizing to Covariate Shift]{Example showing how augmentation and alignment (alignment in the form of normalization and image alignment) can combat covariate shift. From \cite{measuring_domain_shift_for_dl_in_histo}.}
    \label{fig:aug_vs_align}
\end{figure}

Feature alignment techniques for domain generalization attempt to align features across domains to reduce the impact of covariate shift. This may be through the alignment of input data itself or of model-internal representations of the data. This alignment can be done in a "normalization" fashion, decreasing diversity of all domains, or in an "explicit alignment" fashion, which aligns specific, defined domains. Unlike data augmentation, which encourages learning general representations, alignment attempts to remove the need. Figure \ref{fig:aug_vs_align} shows this.

We discuss a simple example of alignment before discussing applications and other techniques from the literature. The idea of alignment is to manipulate a datatpoint $x_i \sim p(x)$ such that it appears to be sampled from another distribution, i.e. $x_i' \sim q(x)$. We would also like to add the stipulation that only irrelevant information is changed; relevant information must be maintained. One long-used method for this, originally from use in natural images \cite{fundamentals_of_digital_image_processing} but also applied to histopathology \cite{histo_matching_histo} is called histogram matching. The idea is to monotonically map an input image's color histogram to another distribution's color histogram such that the darkest color is mapped to the darkest color and the lightest color is mapped to the lightest color. 

Alignment of input data (that is, the images themselves) is widely researched in histopathology, where it is termed stain normalization if done for stain variations and scanner normalization if done for scanner variations, or color normalization more generally \cite{dl_in_dp_survey, a_study_about_color_normalization_for_histo}. In fact, "normalization" in this case is a misnomer, as it includes both normalization and alignment. 

A naive approach to color normalization could be to convert the images to grayscale and to normalize the intensity of the grayscale image. Other approaches may convert the image from RGB-space to a color space that better disentangles intensity and color (e.g. LAB-space), whereupon the color histograms may be aligned. More complex examples include style transfer between domains or GANs to transfer appearance statistics between domains \cite{generalizing_to_unseen_domains_survey}.

These examples all focus on aligning the input data, but as stated before, feature alignment can also encompass alignment of model-internal representations. This is quite similar to input data alignment, and many previous examples have analogous techniques for alignment of internal representations. 

If we have access to data and label pairs from the target domain, we can align joint or conditional distributions, but if we only have unlabeled target data, we can still fix covariate shift by aligning the marginal feature distributions. Remember that, in covariate shift, only $\ptar(x)$ has shifted and that $\ptar(y|x) = \psrc(y|x)$ remains the same. Thus, if we had a way to re-align $\ptar(x)$ to $\psrc(x)$, our joint distribution would be $\ptar(x,y) = \psrc(y|x)\psrc(x)$, the same distribution we trained on. However, this is an overly-optimistic picture; existing marginal alignment techniques are not perfect.

To show the limitations of marginal alignment techniques, we introduce Adaptive Batch Normalization (AdaBN). This popular approach to marginal feature alignment is to align the distribution of the network's layer activations between the source and target set with affine transforms \cite{adabn, nado}. AdaBN leverages the Batch Normalization (BN) layers present in many existing networks, which at test time normally retain the training data's mean and variance parameters-- however, AdaBN replaces those parameters with parameters from the test data, transforming the layer activations in an affine manner. This is found to improve in some cases of covariate shift \cite{limitations_of_post_hoc_feature_alignment}. This technique is post-hoc-- that is, it can be applied to an existing model without the need for pretraining. This removes the need for some other domain generalization techniques to retrain for each new domain, and more importantly, it means AdaBN can be applied to unforseen shifts. 

Unfortunately, simply updating only the BatchNorm parameters is fragile in some covariate shift cases \cite{limitations_of_post_hoc_feature_alignment}. AdaBN performs best on domain shifts that are characterizable as changes in batchnorm statistics; these shifts are simple. This does include style transfer-like domain shifts, as BatchNorm statistic alignment has been shown to be an effective form of style transfer \cite{li_style_transfer_bn}. However, it does not do well on less-uniform shifts. Other, more robust and expressive methods of estimating test domain shift are necessary.

A crop of papers investigate this, proposing methods that use different adaptation techniques to update a larger set of model parameters, such as Tent \cite{tent} and MEMO \cite{memo}. 

Test Entropy Minimization (Tent) is an ad-hoc technique that requires no training examples at test-time. For stability and efficiency, it optimizes affine parameters from linear and low-dimensional operations, but from more than just batchnorm layers. It does this optimization with the goal of minimizing the entropy of a batch of test examples (must be greater than one, the more the better, although there is an adapted technique that works on single images). 

Tent outperforms AdaBN by 5.9\% on ImageNet-C, and performs well specifically on corruptions and some real-world shifts like simulation-to-real discrepancies. However, on undescribable domain shifts, like those found in ImagenetV2 and CIFAR 10.1, this approach does not help.

Minimization with One test point (MEMO) is an ad-hoc technique that only requires one test example and no training examples at test-time. It uses the idea of Test Time Augmentation, when at test-time, a single test point is strongly augmented to create multiple examples which are averaged to create a "marginal output distribution" \cite{TTA}, but takes it further by optimizing all model parameters to minimize the entropy over this averaged output. The intuition behind this is that the model should be invariant to these augmentations, and that confidence is correlated with model correctness, supported by an earlier assumption that decision boundaries in the dataset lie in low density regions of the data space \cite{grandvalet_and_bengio_2005}. It gives 1-8\% improvement in accuracy over non-domain-generalization techniques and sometimes outperforms Tent in some settings, but provides state of the art results specifically in the case of only one test point.


\subsection{Test Time Training (TTT)}
\label{ttt_section}
Test Time Training (TTT) is a feature alignment-based approach to domain generalization that is reactive to domain shifts and requires only one domain for training. This approach was first proposed for natural images \cite{ttt}, and we hope to apply it to histopathology images. The model, trained on a training domain, can be aligned to the test domain by creating a self-supervised learning problem with a test batch (ranging from one to many test examples, with more examples leading to lower variance and a better estimation of the test shift). This self-supervised task can be referred to as the secondary task. 

The hope is that backpropagating on the secondary task when there is some domain shift will allow the main task to perform better as well. However, there are some assumptions that underpin this hope. For one, the secondary task must be correlated with the main task. Because we are not backpropagating with respect to the main task, there is some risk of overfitting to the secondary task and in fact decreasing performance on the primary task. This is also why the training procedure must be changed from optimization solely over the primary task to joint optimization over both tasks. Without this step, there may be some low-hanging fruit for secondary task optimization during test-time, leading to undesirable and unpredictable gradients. Secondly, there must be some measurable domain shift. The original paper shows that this technique provides no improvement when there is no domain shift. 

\begin{figure}
    \centering
    \includegraphics[width=0.66\textwidth]{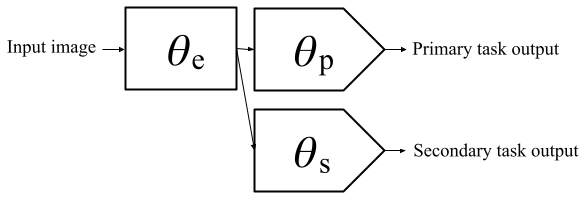}
    \caption[Test Time Training Model Architecture]{Test Time Training Model Architecture.}
    \label{fig:ttt_arch}
\end{figure}

The setup for TTT is applicable to any existing DNN model, with slight architectural modification, visualized in Figure \ref{fig:ttt_arch}. The existing model is conceptually divided into an encoder that maps input images to a latent space and a primary task head that maps the latent space to the output space. From here, a secondary task head is added that maps from the latent space to a secondary output space. 

\subsubsection{TTT vs Other Techniques}
Test Time Training was selected to investigate further in this project. The technique is of interest because it fits especially well to our desired setting. 

TTT is not required access to labels of the target domain, unlike transfer learning, meta learning, and some feature alignment techniques. Similarly, TTT only requires access to one source domain, whereas meta-learning, transfer learning, and other early domain generalization techniques require many.

TTT is not required access to source examples at test time, which is necessary of some other feature alignment techniques. This is beneficial for productization of AI algorithms and the restrictive data permissions in health applications. 

TTT does not require examples of domain shifts or predicted domain shifts at train time. We call this "reactive" adaptation, because we do not need to predict what the shift may look like in advance. This is relatively unique to test-time feature alignment techniques, in contrast to meta learning, transfer learning, GAN-based feature alignment, data augmentation, and self-supervised learning for domain generalization. This is especially important for the invisible domain shifts that arise not from scanner or stain variability, but from causes we do not yet understand.

Comparing TTT to other feature adaptation methods, it modifies more than just a few affine parameters, unlike Tent and AdaBN. This allows for a potentially stronger and more complex adaptation. This may be useful for domain shifts that are less globally uniform.

TTT as a feature adaptation method may get a stronger feature adaptation signal because it uses a secondary task rather than relying on something more naive like test batch statistics like AdaBN or Tent. 

TTT as a feature adaptation method is able to adapt on just one test example, unlike AdaBN and Tent. This is especially critical in the clinical setting. 

\paragraph{Criticisms of TTT}
TTT is not ad-hoc, like some feature adaptation methods: it requires modification of the training procedure (namely, joint training) and selection of a secondary task, and this comes with its own host of problems: how to properly weight primary vs secondary tasks, what task can properly capture the domain shift signal, the question of whether the primary and secondary tasks are adequately correlated, how much to optimize the secondary task at test time, and whether the performance will diminish if there is no domain shift. Additionally, adapting the shared encoder without adapting the primary task head does not guarantee that the primary head will be able to interpret the new encoder output space.





\chapter{Experiment}
\label{experiment_chapter}
We attempt to improve domain shift robustness by applying the Test Time Training (TTT) technique introduced in Section \ref{ttt_section}. We test the two secondary tasks introduced in Asides \ref{RSP_aside} and \ref{SimCLR_aside}: RSP and SimCLR. 

\section{Common Experimental Details}


\subsection{Methods}
\subsubsection{Algorithm}
The model used is a modified version of a proprietary model. The details of the model will only be generally discussed, but the modifications will be described in full. 
The original proprietary model takes input images of size 526x526 with 3 color channels and classifies them as either Normal Tissue, Stroma, or Cancer.

For these experiments, the model is conceptually divided into two modules: an encoder that maps input images to a 512-dimensional latent space, and a primary task head (or, a classifier) that maps from the latent space to the primary task output space.

The modifications to the original model are as follows: a secondary task head is created that maps from the latent space to a secondary output space. This secondary output space is unrelated to the primary task space. The self-supervised secondary task is used to tune the shared encoder, either in joint training at training time, or by itself at test time. 

In the joint training procedure, the supervised loss from the primary task and self-supervised loss from the secondary task are combined and weighted: $$\ell = \left(1-\lambda_\text{s}\right) * \ell_\text{p} + \lambda_\text{s} * \ell_\text{s}.$$ Then, the combined loss $\ell$ is optimized. 

During test-time, the model is only optimized for one step of secondary task backpropagation. In this step, the secondary task head's weights and the shared encoder's weights are changed, while the primary task head's weights are unable to be changed.

We now introduce the implementation details for the two secondary tasks of interest: Resolution Sequence Prediction (RSP) and SimCLR. The original paper's secondary task of image rotation prediction is not used because WSI patches are rotationally ambiguous. We select these two tasks as they show promise in the vanilla self-supervised pretraining setting.

\begin{figure}
    \centering
    \includegraphics[width=\textwidth]{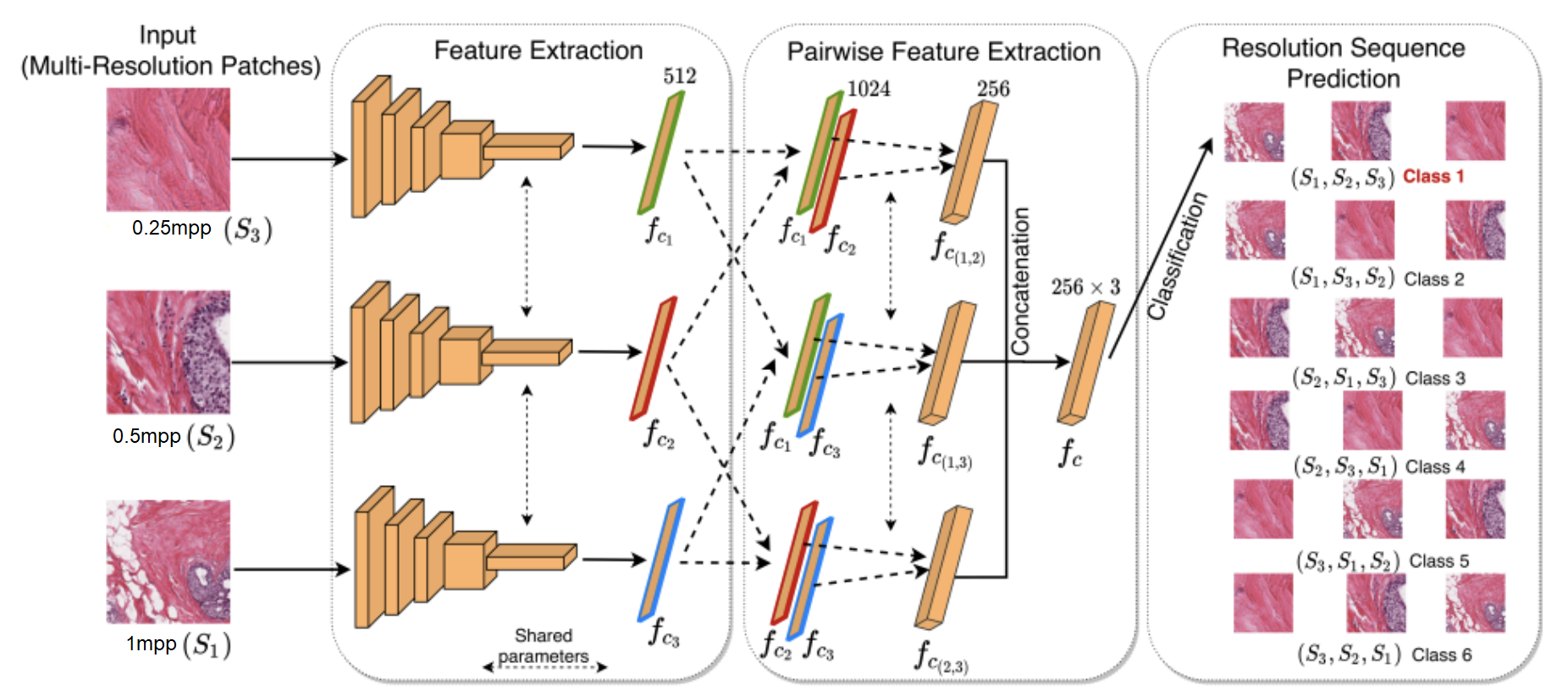}
    \caption[The RSP Secondary Task and Secondary Task Head Architecture]{The RSP secondary task, and the secondadry task head architecture. Adapted from \cite{martel_ssl}.}
    \label{fig:RSP_arch}
\end{figure}

\paragraph{RSP}
The Resolution Sequence Prediction (RSP) task was introduced in \ref{RSP_aside}. Here, we discuss the implementation details. 
The architecture of the shared encoder and main task head remains the same as in the vanilla model. The secondary task head architecture is taken from \cite{martel_ssl} and combines features from all three patches by encoding each patch with the shared encoder, then concatenating the encoded representation for each possible pair of patches, feeding each pair of representations into a two-layer MLP, and finally concatenating this output for all three pairs and feeding it into a two-layer MLP for classificaion. A visualization of this can be seen in \ref{fig:RSP_arch}.

For the joint training procedure of this task, each sample is used in both the primary task and secondary task. For the primary task, just the 0.25mpp resolution patch of the sample is fed through the shared encoder and primary task head. For the secondary task, all three resolution patches are fed through the shared encoder and secondary task head.

\paragraph{SimCLR}
The SimCLR contrastive task was introduced in \ref{SimCLR_aside}. Here, we discuss the implementation details. The architecture of the shared encoder and main task head remains the same as in the vanilla model. The secondary task head architecture is taken from \cite{simclr}-- it is a two-layer MLP that maps from the output space of the shared encoder to the space on which the contrastive loss is applied.

Augmentations used for this task are compositions of $360^{\circ}$ rotations, image flips, strong color jitter, random resize crop and random grayscale. 


\subsubsection{Training Details}
The model is trained under a joint training paradigm \cite{hendrycks_joint_ssl_robustness} on the single training domain. The Adam optimizer is used with standard pytorch initialization and an initial learning rate of 0.001, but the learning rate is multiplied by a gamma factor of 0.5 every 5,000 steps. Every 250 steps, the training step is logged and 120 steps of validation is conducted and logged.

The primary task loss uses cross entropy loss, and basic augmentations are applied with the goal of improving generalization and reducing overfitting, not learning domain invariance. These augmentations are color jitter and image flipping. Any augmentations applied to the primary task are also applied to the secondary task, and applied before the secondary task augmentations.

\paragraph{RSP}
The model is joint trained for 10,000 steps with a batch size of 24. Each batch is stratified with 8 samples of each class. We have to use a smaller than desirable batch size because each sample has three resolutions and each sample is loaded onto GPU in highest resolution. The RSP task loss is cross entropy loss.

\paragraph{SimCLR}
The model is joint trained for 20,000 steps with a batch size of 144. Each batch is stratified with 48 samples of each class. The secondary task uses Normalized Temperature-scaled Cross Entropy Loss, proposed in the original paper \cite{simclr}.

\subsection{Materials}
\subsubsection{Dataset}
The dataset in this project is 373 WSIs from prostate biopsies. The dataset is split into subsets: train (N=232), validation (N=47), test A (N=47), and test B (N=47). The train dataset is used to train model parameters in the standard backpropagation style, the validation dataset is used during training to evaluate training progress and to select a model that should generalize well, the test A dataset is used as the "test" dataset in the original sense, that is, to simulate the deployment setting of a model to tune hyperparameters of test-time-training, and the the test B dataset is used only once after tuning hyperparameters to get an unbiased estimate of the performance improvement from test time training. 

From these WSI datasets, image datasets are created by sampling pathologists' annotations at 0.25 microns per pixel (mpp), creating 526x526 pixel (131.5x131.5 micron) patches for training and inference.

There are two domain shifts being studied in this project: one natural shift arising from scanner variations, and one synthetic shift created by applying gaussian noise to the test set. The scanner variation datasets are constructed by scanning the same set of whole slide images with two different scanners, the Aperio AT2 and the Aperio GT450. Therefore, the datasets include the same slides and the same information.

Due to quirks of sampling, the AT2 and GT450 dataset do not have exactly the same image composition, but they are very close. We describe composition of the AT2 image datasets, with the understanding that the GT450 dataset is very similar. The training set is 553,414 images, class stratified with 200,000 normal tissue images, 200,000 stroma images, and 153,414 cancer images. The validation set is 117,988 images, class stratified with 40,000 normal tissue images, 40,000 stroma images, and 37,988 cancer images. The test A set is 9,396 images, class stratified with 3,333 normal tissue images, 3,333 stroma images, and 2,730 cancer images. The test B set is 9,999 images, class stratified with 3,333 normal tissue images, 3,333 stroma images, and 3,333 cancer images.

Experiment 1 only uses the AT2 dataset, but Experiment 2 uses both the scanner shift and the synthetic shift datasets.


\section{Experiment 1: Tension Between Tasks in Joint Training}
\subsection{Experimental Design}
Prior papers \cite{ttt, tttpp} indicate that a key factor for the success of Test Time Training is a high degree of correlation between the primary task and secondary task. Because, at test time, the secondary task has free rein over the tuning of the shared encoder parameters, this argument makes sense. However, the two papers make no further discussion on this-- it is unanswered whether this task is easy in practice. This experiment investigates the level of difficulty in practice to find an appropriate secondary task that balances well with the primary task during joint training only.

We also wish to see whether these secondary tasks work "well" for joint training-- they should improve performance on the primary task when introduced.

\subsubsection{Hypothesis}
We hypothesized that taking secondary tasks from SSL pretraining tasks for histopathology would prove to be successful for joint training. Based on joint training literature \cite{hendrycks_joint_ssl_robustness}, joint training (with both proposed secondary tasks: RSP and SimCLR) should provide better results than vanilla training.

We also further investigate the RSP task, seeing if representations learned from pretraining on it are useful when fine-tuning on the main task. We hypothesize that a model pretrained on the RSP task and fine-tuned on the main task will perform better than a model randomly initialized and trained on the main task.

\subsubsection{Methods}
This is a three-pronged experiment attempting to identify whether joint training (a prerequisite for TTT) is feasible with the two proposed tasks. For \expOnesOne and \expOnesThree, the RSP setup is used, and for \expOnesTwo, the SimCLR setup is used. 

\paragraph{\expOnesOne}
This experiment conducts joint training with the RSP secondary task while varying the secondary weighting parameter $\lambda_\text{s}$. A hyperparameter search of $\lambda_\text{s}$ is conducted on the logarithmic scale between $1 \times 10^{-4}$ and $1 \times 10^{-1}$, finding that $\lambda_\text{s} = 1 \times 10^{-2}$ is the optimal value. We record the training and validation loss curves through $10,000$ steps with a batch size of 24.

\paragraph{\expOnesThree}
This experiment investigates the RSP task in a self-supervised pretraining setting instead of a joint training setting. It tests whether representations learned from the RSP task are preserved after fine-tuning to the main task. We compare a model that was pretrained with the RSP task for $10,000$ steps to a model that is randomly initialized. In this experiment, $\lambda_\text{s} = 0$. We record the training and validation loss curves through $10,000$ steps with a batch size of 24 examples.

\paragraph{\expOnesTwo}
This experiment is similar to \expOnesOne but with the SimCLR secondary task setup instead of RSP. The $\lambda_\text{s}$ parameter is searched between $1 \times 10^{-3}$ and $5 \times 10^{-2}$, again finding the optimal $\lambda_\text{s} = 1 \times 10^{-2}$. We record the training and validation loss curves through $20,000$ steps with a batch size of 144.


\subsection{Results}
\paragraph{\expOnesOne}

\begin{figure}
    \centering
    \includegraphics[width=\textwidth]{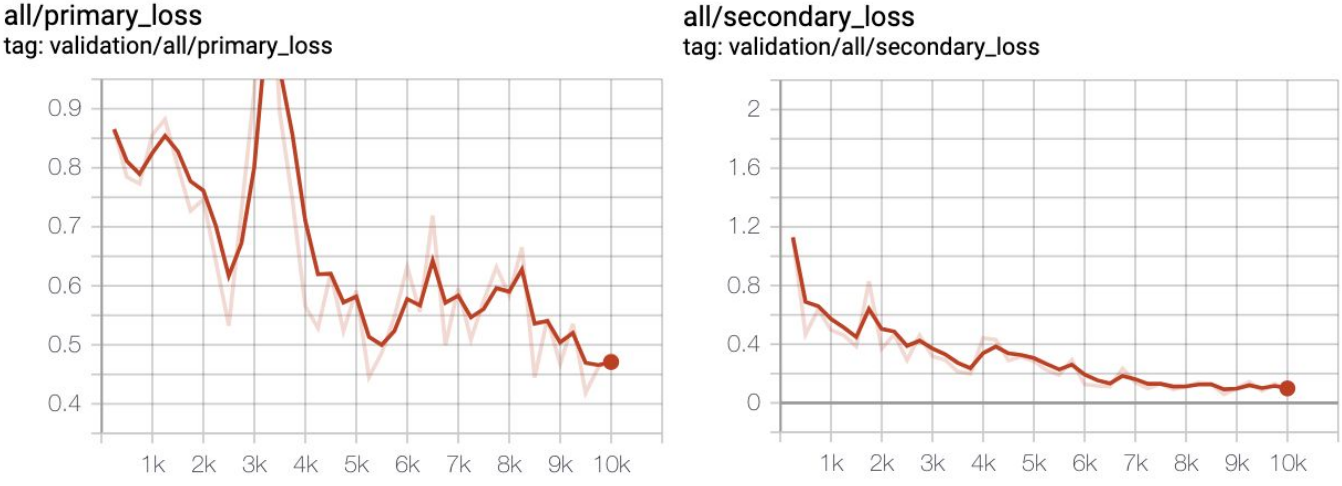}
    \caption[\expOnesOne: Best Joint-trained Model Loss Curves]{The primary loss and secondary loss curves for the best joint-trained model, evaluated on the validation set.}
    \label{fig:results_tboard_1_1}
\end{figure}

The vanilla model (trained without a secondary task) achieves validation primary loss of 0.445, while the best joint-trained model (trained with the secondary task at $\lambda_\text{s} = 1 \times 10^{-2}$) outperforms it and reaches validation loss of 0.419. The secondary loss of the vanilla model is 1.796 (the vanilla model is not incentivized to improve this metric at all), and the secondary loss of the best joint-trained model is 0.0593. We record the secondary loss for both to ensure that the best joint-trained model actually learns the secondary task. Results for other $\lambda_\text{s}$ are shown in Table \ref{tbl:exp11}. The validation primary loss and validation secondary loss curves for the best joint-trained model are shown in Figure \ref{fig:results_tboard_1_1}.

\begin{table}
\centering
\begin{tabular}{l|l|l}
                     & Validation loss (primary task) & Validation loss (secondary task) \\ \hline
Vanilla: $\lambda_\text{s}=0$ & 0.445                 & 1.796                 \\
RSP: $\lambda_\text{s}=0.0001$
     & 0.416                 & 1.005                 \\
RSP: $\lambda_\text{s}=0.001$   & 0.414                 & 0.853                 \\
RSP: $\lambda_\text{s}=0.01$ & 0.419                 & 0.059                 \\
RSP: $\lambda_\text{s}=0.1$    & 1.025                 & 0.000    
\end{tabular}
\caption[\expOnesOne: Best Primary Loss and Associated Secondary Loss for Each Training Run]{The best primary task validation loss and concurrent secondary task validation loss for each training run in the hyperparameter search for $\lambda_\text{s}$ in \expOnesOne.}
\label{tbl:exp11}
\end{table}

\paragraph{\expOnesThree}
The best validation loss of the primary task in the random weight initialization case is 0.445, whereas it is worse in the RSP pretrained case at 0.465. Again, we monitor the secondary loss for both. The validation loss of the secondary task in the random weight initialization case is always between 1.796 and 1.799; it remains essentially constant (note: this makes sense, as $\lambda_\text{s} = 0$).  The validation loss of the secondary task in the RSP pretrained case, however, begins at a low of $1 \times 10^{-7}$ and reaches 2.53 when the best primary loss is attained.

\begin{figure}
    \centering
    \includegraphics[width=0.5\textwidth]{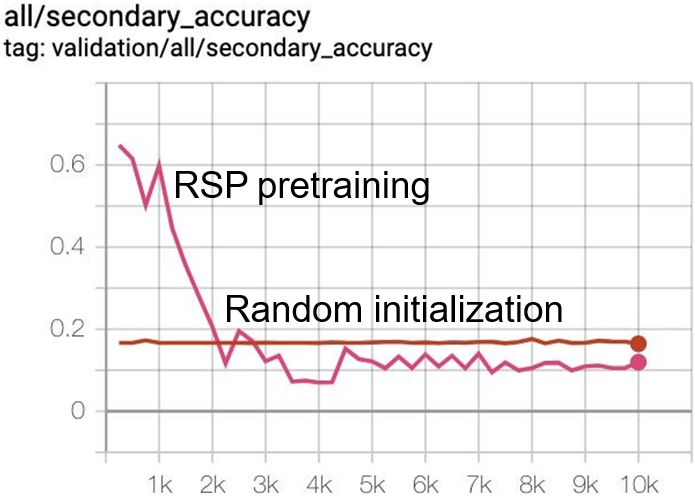}
    \caption[\expOnesThree: Difference in Secondary Task Accuracy Curve With and Without RSP pretraining]{Validation secondary task accuracy for RSP pretrained and randomly initialized network.}
    \label{fig:NEWresults_tboard_1_3}
\end{figure}

The validation secondary task accuracy curves for these two runs are seen in \ref{fig:NEWresults_tboard_1_3}. 

\paragraph{\expOnesTwo}
The vanilla model (trained without a secondary task) achieves validation primary loss of 0.311, while the best joint-trained model (trained with the secondary task at $\lambda_\text{s} = 1 \times 10^{-2}$) outperforms it and reaches validation loss of 0.280. The secondary loss of the vanilla model is 10.42 (the vanilla model is not incentivized to improve this metric at all), and the secondary loss of the best joint-trained model is 9.86. We record the secondary loss for both to ensure that the best joint-trained model is learning the secondary task. Results for other $\lambda_\text{s}$ are shown in Table \ref{tbl:exp12}.

\begin{table}
\centering
\begin{tabular}{l|l|l}
                                     & Validation loss (primary task) & Validation loss (secondary task)  \\ 
\hline
Vanilla: $\lambda_\text{s}=0$        & 0.311~ ~              & 10.42~ ~              \\
RSP: $\lambda_\text{s}=0.001$   & 0.312                 & 10.39                 \\
RSP: $\lambda_\text{s}=0.01$  & 0.280~ ~              & 9.86                  \\
RSP: $\lambda_\text{s}=0.05$   & 0.367                 & 9.73                 
\end{tabular}
\caption[\expOnesTwo: Best Primary Loss and Associated Secondary Loss for Each Training Run]{The best primary task validation loss and concurrent secondary task validation loss for each training run in the hyperparameter search for $\lambda_\text{s}$ in \expOnesTwo.}
\label{tbl:exp12}
\end{table}


\subsection{Discussion}
\paragraph{\expOnesOne}
Joint training with RSP works "well"-- as intended, the performance of the primary task is better with the joint training than without it. 

However, choosing the proper task weighting parameter $\lambda_\text{s}$ is difficult. If there is too much weight on the secondary task, the primary task suffers. In the $\lambda_\text{s} = 0.1$ case, the primary task generalizes poorly to the validation set while the secondary task generalizes nearly perfectly. This suggests that the model has learned flimsy representations for the primary task due to the optimization direction of the secondary task overpowering the direction of the primary task during training. This suggests that the main task and the secondary task are in opposition, suggesting that this task may not be appropriate for TTT. 

Even though the primary loss does benefit from joint training with RSP, notice that the secondary loss has already optimized to a plateau halfway through the experiment while the primary loss does not reach a minimum, even with twice as many steps. And in the $\lambda_\text{s} = 0.1$ case, the secondary task is optimized to less than 1/10th than in the $\lambda_\text{s} = 0.01$ case. This may suggest that the secondary task is easier to learn than the primary task. 

Althogh we observe good joint training behavior for RSP, the ease with which the secondary task overpowers the first task is somewhat concerning. \expOnesThree investigates further whether RSP is a good fit for TTT.


\paragraph{\expOnesThree}
The SSL pretrained run sees a decrease in performance over the random initialization run with respect to the primary task. This may indicate that the representations learned for the secondary task were not transferable to the primary task. Observing the secondary loss for the random initialization run tells us that random guessing on the secondary task should give a loss around $1.775$. However, the representations produced by the shared encoder stray far from the representations expected by the trained secondary task head in the pretrained run, producing a secondary loss higher than the random guessing baseline despite having some (pre-trained) supervision on the secondary task. This can also be seen in \ref{fig:NEWresults_tboard_1_3}-- the random initialization model achieves accuracy equivalent to random guessing, but the pretrained model does worse than this. The pretrained model's secondary task head is left untouched during fine-tuning, while the shared encoder changes, resulting in this mismatch of encoding supports.

These results suggest that the main task and the secondary task are in opposition, discrediting the use of RSP for TTT. 

\paragraph{\expOnesTwo}
Like the RSP experiments, properly setting a  $\lambda_\text{s}$ is nontrivial.

Compared to the RSP task, the SimCLR task helps decrease the validation loss to a greater degree. Although the baseline for the experiments are different due to continued development, the gap is larger for SimCLR than it is for RSP. The SimCLR task also does not appear to be "too easy" to learn in relation to the primary task: both losses decrease through the entire training procedure. 

The results of this experiment, showing that SimCLR works well for joint training, suggest that SimCLR will also be a good fit for Test Time Training.

\subsection{Conclusion}
The proper balance for $\lambda_\text{s}$ is difficult to find and may vary for different tasks and datasets. When the tasks are unbalanced, performance on the primary task can be greatly diminished. 

RSP seems to be less appropriate for TTT than SimCLR. RSP may be too easy a task, or it may be too in-opposition to the primary task. SimCLR seems to be more appropriate and is not learned too quickly compared to the primary task.

\section{Experiment 2: Tension Between Tasks in Test-Time Adaptation}
\subsection{Experimental Design}
Results from Experiment 1 show that our model has been effectively joint trained on the primary task and the SimCLR secondary task. This model should be ready for test-time training.

\subsubsection{Hypothesis}
If we use a model that has learned both tasks through joint training, we hope that test-time training on the secondary task will yield improved results on the primary task if the domain has shifted. An additional hope is that test-time training does not hurt performance if the domain has not shifted.

\subsubsection{Methods}
We use the SimCLR secondary task model with $\lambda_\text{s}=0.01$ trained in Experiment 1, \expOnesTwo. We do test time training with this model on three settings: two with test-time domain shifts and one with no domain shift. The training domain is AT2 scanned WSIs, and the two domain shifts are an in-the-wild scanner shift created by scanning the test set of slides with a GT450 scanner, and a synthesized domain shift created by adding Gaussian noise to an AT2 test set. We test an in-the-wild scanner shift because it is most similar to what we see in real life, and we test the Gaussian noise corrupted dataset because it is the setting in which TTT does best in the original paper.


\subsection{Results}
\begin{figure}
\centering
\begin{subfigure}{.5\textwidth}
  \centering
  \includegraphics[width=\linewidth]{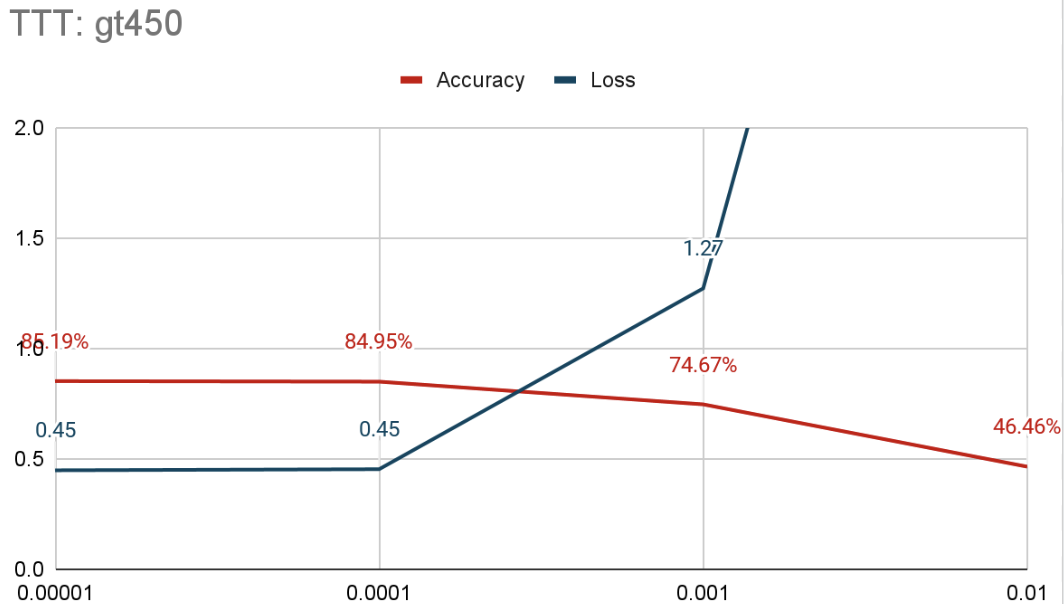}
  \caption{Varying the TTT step size for a scanner domain shift.}
  \label{fig:results2_1}
\end{subfigure}%
\begin{subfigure}{.5\textwidth}
  \centering
  \includegraphics[width=\linewidth]{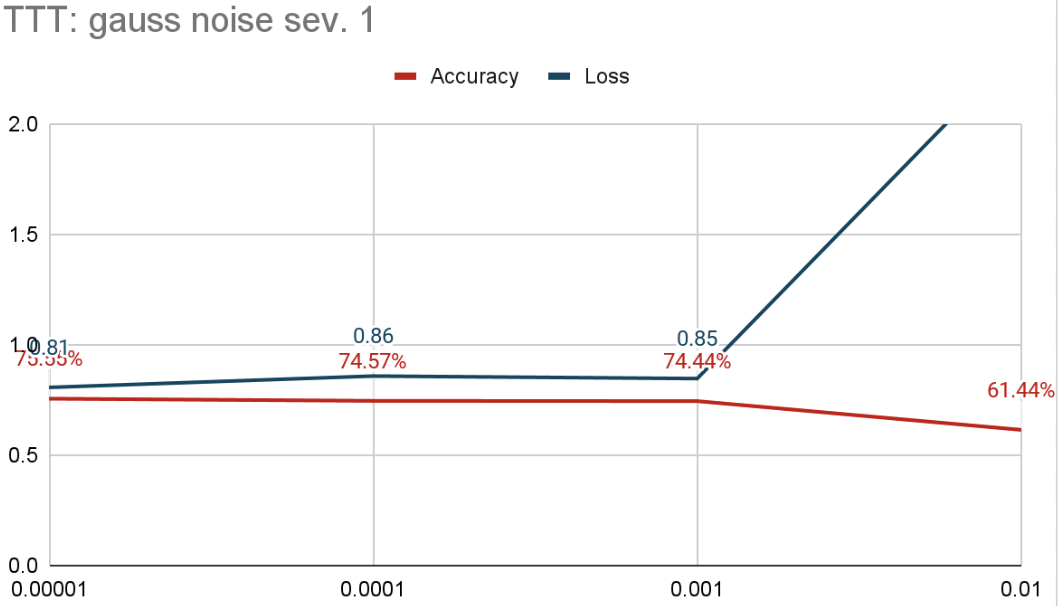}
  \caption{Varying the TTT step size for a simulated domain shift, generated by adding gaussian noise to the test set.}
  \label{fig:results2_2}
\end{subfigure}
\begin{subfigure}{.5\textwidth}
  \centering
  \includegraphics[width=\linewidth]{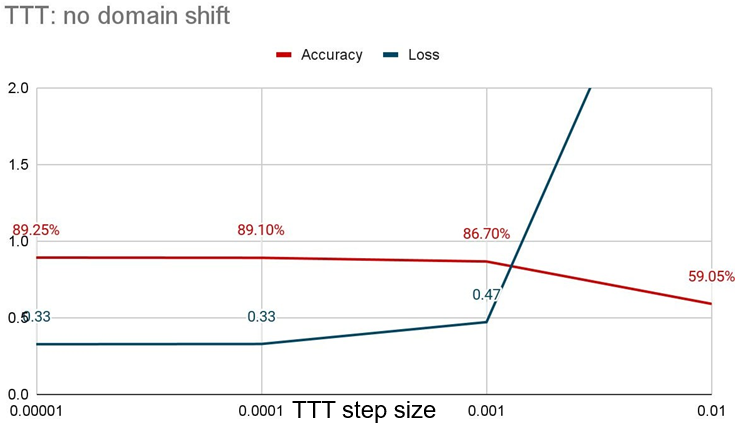}
  \caption{Varying the TTT step size for no domain shift.}
  \label{fig:results2_1}
\end{subfigure}%
\caption[Test Set Accuracy and Loss vs TTT Step Size for our Three Settings.]{Test set accuracy and loss vs. TTT step size for our three settings. In case color is not visible, accuracy is labeled with percentage numbers.}
\label{fig:results2}
\end{figure}

Loss and accuracy on a class-balanced test set for the three settings, with varying step sizes for TTT can be seen in \ref{fig:results2}. 

\subsection{Discussion}
The TTT technique is supposed to shine when there is a domain shift at test time. However, we see that the loss on the primary task increases as a large step size is used, and the performance stays steady as the step size gets asymptotically smaller. The TTT procedure has too little effect if the step size is too small, or it moves performance in the wrong direction if a proper step size is used. We see the same problem in the setting without domain shift.

Despite our original model being joint trained to achieve good optimization on both tasks, the agreement between both tasks is not sufficient in this case to improve the primary task by optimizing the secondary task.

\subsection{Conclusion}
In both a natural case of domain shift and in a controlled synthetic case that performed favorably in the original paper, we see no benefit from TTT. The optimization of the SimCLR secondary task is not correlated enough with the optimization of the primary task for TTT to work as intended. A consideration is that perhaps SimCLR functions better as a coarse pretraining task but not for fine-grained optimization.

An existing paper based on test time training suggests a new form of regularization to ensure that the TTT optimization does not cause the shared encoder weights to stray too far from their initial values. Implementing this improves results by a lot on the synthetic domain shift and marginally on the real domain shift, but their ablation shows that TTT with a contrastive learning task (equivalent to our method) should perform well \cite{tttpp}. 

Future work to investigate this would begin by determining whether there exists any signal from the secondary task optimization that is useful for feature alignment in a domain shift. It may be beneficial to constrain weights from varying too much; or beneficical to only change some specific weights (similar to Tent and AdaBN, which only change some affine parameters \cite{tent, adabn}). Modifying all parameters may be overfitting to the secondary task and destroying meaningful representations learned during joint training.




\chapter{Conclusion}
\label{conclusion_chapter}
There is a need for better domain generalization techniques in AI-based digital pathology. Different domains arise commonly in digital pathology and it is important for models to generalize to different domains without annotations on the new domain and without knowledge of what the domain shift may look like. 

New research in domain generalization, specifically through feature alignment, proves quite interesting for these reasons. The technique of focus for this project is Test Time Training, a technique for many-parameter test-time feature alignment through self-supervised task labels. It is reactive, rather than predictive, of domain shifts; it requires only one source domain, with no access to it at test time; and it is capable of modeling complex domain shifts.

Despite the claimed benefits, our experiments on this technique show the deficiencies of this approach for our setting. This approach is not ad-hoc: it requires modification of the training procedure, and additional tuning to find an appropriate secondary task and proper weighting for them. Even with an appropriate task and appropriate weighting, Test Time Training did not provide us with increased domain generalization performance in this setting.

Future work investigating the efficacy of TTT in this area should explore the secondary task optimization at test time to see if there is any valuable information for feature alignment. It may be beneficial to constrain weights from varying too much; or beneficical to only change some specific weights (similar to Tent and AdaBN \cite{tent, adabn}).

For future work in domain generalization in this area, it may be useful to investigate truly ad-hoc domain generalization techniques as well. The literature shows other adaptation techniques peforming comparably to Test Time Training, and these techniques remove the headache of changing the training procedure.

In conclusion, although this project shows negative results for Test Time Training on domain generalization, it motivates the need for reactive, not predictive, domain generalization techniques in computational histopathology. 


\bibliography{thesis}

\appendix

\end{document}